\documentclass[iop]{emulateapj}
\usepackage{amsmath}

\shorttitle{A Relation between the WNM and WIM in the CGPS}
\shortauthors{Foster et al.}

\begin{document}

\title{A Relation Between the Warm Neutral and Ionized Media
Observed in the Canadian Galactic Plane Survey}

\author{T. Foster\altaffilmark{1,2}, R. Kothes\altaffilmark{2}, 
   and J. C. Brown\altaffilmark{3}}

\altaffiltext{1}{Department of Physics and Astronomy, Brandon University, 
              270 18th Street, Brandon, Manitoba, R7A 6A9, Canada}

\altaffiltext{2}{National Research Council Canada,
              Dominion Radio Astrophysical Observatory,
              P.O. Box 248, Penticton, British Columbia, V2A 6J9, Canada}

\altaffiltext{3}{Department of Physics and Astronomy, University of Calgary, 
              2500 University Dr. N.W., Calgary, Alberta, T2N 1N4, Canada}

\email{Tyler.Foster@nrc-cnrc.gc.ca}

\begin{abstract}
We report on a comparison between 21~cm rotation measure (RM) and the 
optically-thin atomic hydrogen column density 
(N$_{\textsc{HI}}\left(\tau\rightarrow 0\right)$) measured towards unresolved 
extragalactic sources in the Galactic plane of the northern sky.
\ion{H}{1} column densities integrated to the Galactic edge are measured 
immediately surrounding each of nearly 2000 sources in 1-arcminute 21~cm line 
data, and are compared to RMs observed from polarized emission of each source. 
RM data are binned in column-density bins 4$\times$10$^{20}$~cm$^{-2}$ wide, 
and one observes a strong relationship between the number of hydrogen atoms in 
a 1~cm$^2$ column through the plane and the mean RM along the same 
line-of-sight and path length. The relationship is linear over one order of 
magnitude (from 0.8-14$\times$10$^{21}$~atoms~cm$^{-2}$) of column densities, 
with a constant 
RM/N$_{\textsc{HI}}\sim-$23.2$\pm$2.3~rad~m$^{-2}$/10$^{21}$~atoms~cm$^{-2}$, 
and a positive RM of 45.0$\pm$13.8~rad~m$^{-2}$ in the presence of no atomic 
hydrogen. This slope is used to calculate a mean volume-averaged  
magnetic field in the 2$^{\footnotesize{\textrm{nd}}}$ quadrant of 
$\left<B_{\parallel}\right>\sim$1.0$\pm$0.1~$\mu$G directed away from the Sun, 
assuming an ionization fraction of 8\% (consistent with the WNM). %It also suggests an 
%ionization fraction for the Warm Neutral Medium of $\left<x_e\right>\geq$6.3\% 
%in the Galactic plane, using as an upper-limit to the line-of-sight field a 
The remarkable consistency between this field and $\left<B\right>=$1.2~$\mu$G 
found with the same RM sources and a Galactic model of dispersion measures
suggests that electrons in the partially ionized WNM are mainly responsible for
pulsar dispersion measures, and thus the partially-ionized WNM is the dominant 
form of the magneto-ionic interstellar medium.
%The data also tentatively suggest a flattening-out at high rotation 
%measures $|RM|\gtrsim$300~rad~m$^{-2}$ where no extra atomic column density is
%present, probably indicating lines-of-sight that pass through completely 
%ionized portions of the interstellar medium (ISM).

\end{abstract}

\keywords{ISM: magnetic fields --- ISM: atoms --- techniques: polarimetric}

\section{Introduction}
The Milky Way Interstellar Medium (ISM) harbours gas in a variety of 
temperature and density regimes, all coexisting in an uneasy equilibrium 
readily disrupted by large scale forces (e.g. gravitational) or smaller scale 
ones (e.g. stellar winds, shocks). Two of these regimes are the Warm-Neutral 
(WNM) and Warm-Ionized Medium (WIM). The Galactic magnetic field (GMF) couples 
to the ionized component of the ISM creating the so-called Magneto-Ionic Medium 
(MIM), and plays an essential but not well-understood role in the mixing of the 
neutral and ionic ISM phases. Because of the partial ionization by cosmic rays 
and X rays the fractional ionization of the WNM, $x_e$, is non-zero. However it 
remains very much an unreliably measured parameter, and is key to ISM studies 
\citep[see review;][]{kulk88}. Nonetheless, this partial ionization blurs the 
boundary between the WNM and the WIM/MIM somewhat, and hence the field may be 
expected to couple with the WNM as well. To what degree is a matter of debate.

The MIM and WNM can be linked by two observables: rotation measure (RM) and 
atomic hydrogen column density (N$_{\textsc{HI}}$). Each can be respectively 
written as path integrals of spatial densities of electrons ($n_e$) and of
atoms ($n_{\textsc{HI}}$): %; both in units of cm$^{-3}$):
%\begin{align*}
\begin{equation}
\frac{\textrm{RM}}{\left(\textrm{rad~m}^{-2}\right)}
=0.81\left<B_{\parallel}\right>\times\textrm{DM}=0.81\frac{\left<B_{\parallel}\right>}{\left(\mu\textrm{G}\right)}\int
\frac{n_e}{\left(\textrm{cm}^{-3}\right)}\frac{dr}{\left(\textrm{pc}\right)} \\
\end{equation}
\begin{equation}
\label{NHI}
\frac{\textrm{N}_{\textsc{HI}}}{\left(\textrm{cm}^{-2}\right)}
=1.82\times10^{18}\sum_{i}\frac{T_{\textrm{B}}\left(v_i\right)}{\left(\textsc{K}\right)}\frac{\Delta
v}{\left(\textrm{km~s}^{-1}\right)}=\int\frac{n_{\textsc{HI}}}{\left(\textrm{cm}^{-3}\right)}\frac{dr}{\left(\textrm{cm}\right)}
%\end{align*}
\end{equation}
where the path integrals are taken over the entire observable path through the 
Galactic disc, and the magnetic field $B_{\parallel}$ is the line-of-sight 
(LOS) average value (indicated by brackets $\left<~\right>$) and is assumed
uncorrelated with the electron density $n_e$. The ratio between RM and
N$_{\textsc{HI}}$ is then related to the ionization fraction
\begin{equation}\label{xe}
x_e=n_e/\left(n_p+n_{\textsc{HI}}\right)\simeq1/\left(1+n_{\textsc{HI}}/n_e\right)
\end{equation}
by
\begin{equation}\label{ratio}
\frac{\textrm{RM}}{\textrm{N}_{\textsc{HI}}}\left(\frac{\textrm{rad~m}^{-2}}{10^{21}~\textrm{atoms~cm}^{-2}}\right)\simeq 2.632\times10^2\left<B_{\parallel}\right>
\left(\frac{1}{\left<x_e\right>}-1\right)^{-1}
\end{equation}
%which has units of radians of rotation per number of atoms in the column. 
where $\left<x_e\right>$ is the path-averaged ionization fraction, related by 
Eqn.~\ref{xe} to the ratio of path-averaged atomic and electron densities, i.e. 
$\left<n_{\textsc{HI}}\right>/\left<n_e\right>
=\left(\left<n_{\textsc{HI}}\right>\int dr\right)/\left(\left<n_e\right>\int dr\right) 
= \int n_{\textsc{HI}}dr/\int n_edr$. As long as both are observed through the 
same length column, RM/N$_{\textsc{HI}}$ is independent of path length, and 
Eqn.~\ref{ratio} is true for any column length. Equation~\ref{ratio} links the 
WNM and MIM through the ionization fraction, and is an alternative expression 
to the usual ratio of rotation-to-dispersion measure RM/DM for estimating 
$\left<B_{\parallel}\right>$ (see Sec.~\ref{B-est}). 

An order of magnitude estimate of the ratio RM/N$_{\textsc{HI}}$ can be made
with currently accepted estimates for the LOS magnetic field strength in the 
ISM. $\left<B\right>\approx-$1 to $-$2~$\mu$G is found by various authors for 
the uniform field \citep[e.g.][]{sun08,vane11} near the Sun (where the
negative sign indicates the field is directed away from the receiver),
thus the LOS component $\left<B_{\parallel}\right>$ is at most these values or 
smaller. For the ionization fraction in the WNM, \citet{jenk13} find 8\% at a 
canonical WNM density of $n_{\textsc{H}}=$0.5~cm$^{-3}$ within a few hundred pc 
of the Sun, whereas \citet{wolf95} obtain 2\% for the WNM from a 
2-phase model calculation. We use a mean of $x_e=$5\% here for illustration; 
from Eqn.~\ref{ratio} then one might expect to find ratios in the range of 
$-$15 to $-$30~rad~m$^{-2}$ per 10$^{21}$~atoms~cm$^{-2}$.

\section{Observations \& Method}\label{obs}
To measure the ratio RM/N$_{\textsc{HI}}$ in Eqn.~\ref{ratio} we use RMs of 
1970 extragalactic sources, calculated from CGPS \citep[]{tayl03} Stokes Q \& U 
data in four 7.5~MHz-wide sub-bands observed around the 1420~MHz line. The CGPS 
is an interferometric survey that extends from the first to the third quadrants 
of the Galactic plane, covering 52$\degr \leq \ell \leq$193$\degr$, and 
$-$3.5$\degr \leq b \leq +$5.5$\degr$ in latitude, with a small extension up to 
$b=+$18$\degr$ in the range 99$\degr \leq\ell\leq$118$\degr$\footnote{The 
complete CGPS 21~cm line, continuum and polarisation data are available at the 
CADC; http://cadc.hia.nrc.ca}. The method used to extract the RMs is described 
in \citet{brow03}. We assume that there is little Faraday rotation from both 
the inter-galactic medium and from within the source itself, relative to the 
path-integrated Galactic disc. Typical uncertainties in RM are in the range of 
5-20\%. The catalogue of CGPS RMs, uncertainties and statistics is presented in 
Brown et al. (2013, in prep.). The high density of RMs in this 
catalogue provides an unmatched ability to trace large-scale Galactic 
magnetism. For example, \citet{brow03} demonstrate that while some CGPS RMs 
suffer from random rotation due to \ion{H}{2} regions along the LOS, removing 
RMs most correlated with Emission Measure (EM$\propto n_e^2$, traced by 
H$\alpha$ emission towards them) does not change the distribution of the sample 
nor the mean RM in any significant way. The catalogue we use here includes the 
sources in \citet{brow03} but is more extended; nonetheless, while some RMs in 
our sample are undoubtedly affected by \ion{H}{2} regions, the option to 
average RMs within a narrow range of longitudes provides the ability to extract 
the large-scale structure of the field, and to trace the MIM on the large-scale 
\citep[also demonstrated by][for southern-sky RMs]{brow07}.

Integrated atomic hydrogen column densities towards each background source are
measured from CGPS 21~cm \ion{H}{1} line data, which has essentially the same 
elliptical synthesized beam as the 21~cm continuum 
(FWHM$\sim$1$\arcmin\times$1$\arcmin\textrm{cosec}\delta$), and a spectral 
resolution of 1.3~km~s$^{-1}$. Each velocity channel is 
$\Delta v=$0.824~km~s$^{-1}$ wide. The mean \ion{H}{1} brightness temperature
$T_{\textrm{B}}$ (Kelvin) in the $i$-th channel is measured in an annulus 
around each source with an inner radius of 1~FWHM in each of the major and 
minor axes of the elliptical beam, and oriented in the same direction as the 
beam. The annulus has a width of 36$\arcsec$ in both dimensions. The observed 
column density is accumulated as in Eqn.~\ref{NHI} over 256 velocity channels.
%The column density of the \ion{H}{1} emission in each channel is thus 
%N$_{\textsc{HI}}\left(v\right)=$0.0018$T_{\textrm{B}}\left(v\right)\Delta v\times$10$^{21}$~atoms~cm$^{-2}$. 
The sum through all velocity channels records the column density through the 
entire Galactic disc, which is the same path length over which RMs are 
accumulated. Equation~\ref{NHI} is appropriate for warm ($T_{k}\gtrsim$200~K) 
and hence optically-thin ($\tau\rightarrow$0) atomic H, which is the main 
constituent of the WNM and the phase which would most likely couple with the 
WIM/MIM. The additional advantage of integrating the \ion{H}{1} 
emission is that our column densities emphasize the 
smooth large-scale Galactic \ion{H}{1} distribution of the WNM, as opposed to 
\ion{H}{1} absorption which shows very local, dense (and optically-thick) 
neutral hydrogen clouds that mainly are found associated with discrete features 
like young \ion{H}{2} regions \citep[e.g., stellar wind bubbles and expanding
shocked shells,][respectively]{fost04,koth02} and spiral arm 
shock-fronts \citep[][]{gibs05}. These denser discrete clouds are not included 
in the column density integrated from \ion{H}{1} emission only.

The uncertainty in each value of N$_{\textsc{HI}}$ is estimated from the 
%\textrm{\footnotesize{H{\small\rmfamily\@Roman{I}}}}}$ %\small\rmfamily\@Roman{#2}
standard deviation of brightness temperatures within the annulus in each 
\ion{H}{1} channel. This value is typically $\pm$2-3~K per channel, and these 
are summed in quadrature to obtain an estimate for our error in 
N$_{\textsc{HI}}$.  
%will be dominated by deviations from the assumption that the column density 
%directly \textit{towards} the source is the same as that taken immediately 
%\textit{around} it. To estimate this, our code also looks 0.5 degrees off the 
%source in a random direction, and calculates N$_{\textsc{HI}}$ twice, from 
%pixels within the annulus and from the central pixels not within the annulus. 
%The distribution of relative (percent) differences between the two column 
%densities calculated this way shows typically the match is good to $\pm$XX\%, 
%which we will take as our error in N$_{\textsc{HI}}$.

\section{Results}\label{results}
\begin{figure*}[!ht]
%\centerline{
\includegraphics[width=1\textwidth,clip=]{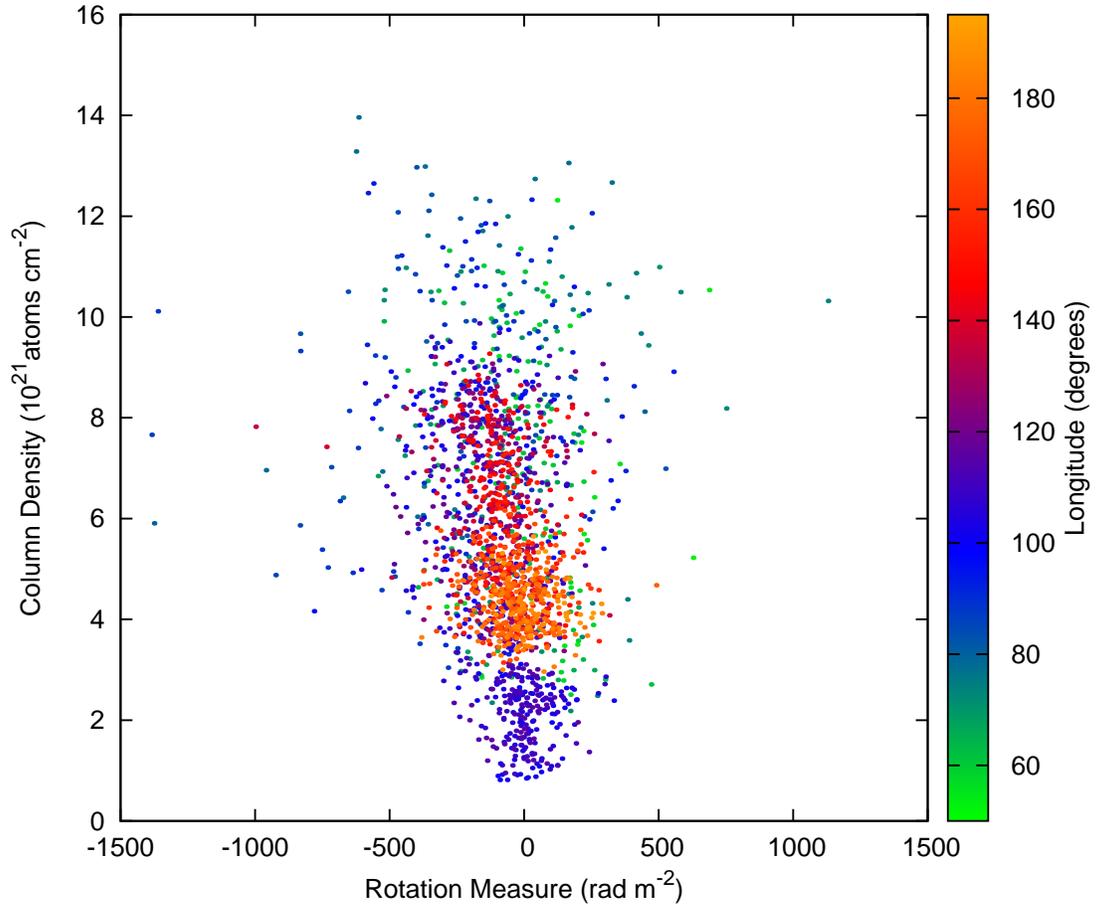}
%}
\caption{Atomic hydrogen column density versus Rotation Measure for 1970
sources (directions) in the CGPS. Data points are coloured by Galactic
longitude (in degrees, colour-bar at right). Error bars are omitted for
clarity.
}
\label{plot1}
\end{figure*}

Figure~\ref{plot1} shows a plot of \ion{H}{1} column density 
N$_{\textsc{HI}}\left(\tau\rightarrow 0\right)$ (units of 
10$^{21}$~atoms~cm$^{-2}$) versus RM (rad~m$^{-2}$). An underlying trend of 
increasing magnitude of RM (negative values) with increasing column densities 
is readily seen, and the scatter about a mean RM for a given column 
N$_{\textsc{HI}}$ does not overwhelm this trend. We see RMs on either side of 
zero, suggesting that a random component to the field plays a significant role 
in the scatter. The mean RM through the CGPS is negative, reflecting the LOS
projection of the GMF in this area of the Milky Way which points away from the 
Sun. The scatter in RM for a given column can come from three broad sources: i) 
scatter caused by a small-scale {}``random'' component related to the 
large-scale GMF, ii) scatter from large-scale magnetic field reversals along 
the LOS, and iii) scatter from smaller scale {}``anomalous'' regions with a 
magnetic field independent of the GMF and/or electron densities different from 
the bulk MIM (such as \ion{H}{2} regions).

\begin{figure*}[!ht]
%\hspace{-2.5cm}
%\includegraphics[scale=1]{plot4a.eps}\hspace{-2.5cm}
\centerline{\includegraphics[scale=0.53]{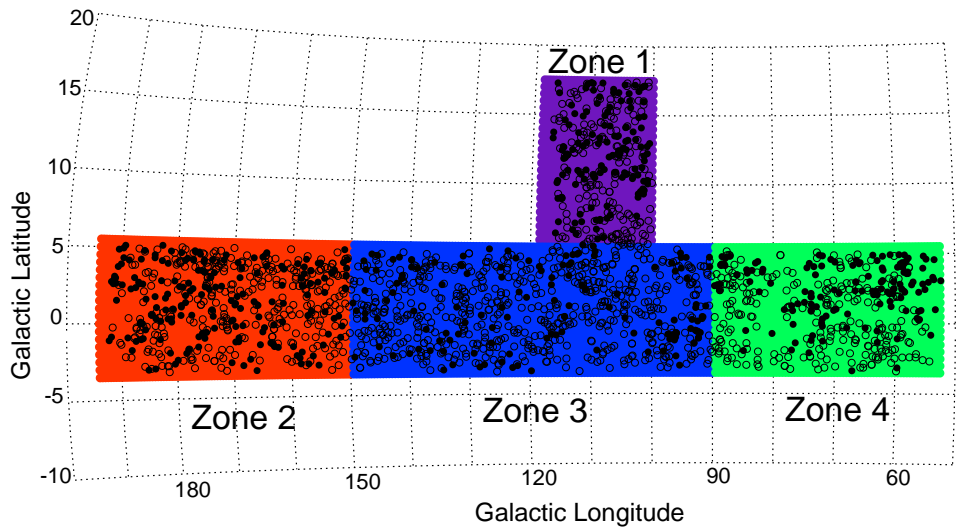}}
\caption{Distribution of all RM sources across the Galactic plane observed
for the CGPS. Zones marked are discussed in Sec.~\ref{results}. Filled
circles indicate RM$>$0, and open circles indicate RM$<$0.}
\label{distro}
\end{figure*}

The distribution and sign ($+,~-$) of the RMs across the CGPS is shown in 
Figure~\ref{distro}, with longitude and latitude zones coloured to roughly 
match the colour bar in Figure~\ref{plot1}. Generally in the plane 
($\left|b\right|<$5$\degr$), lower N$_{\textsc{HI}}$ are observed towards 
higher longitudes. We can divide Figure~\ref{plot1} into four column density 
zones show in Figure~\ref{distro}, from lowest to highest: 1) high-latitude 
sources with +5$\degr<b<+$18$\degr$ which all have measured column densities 
lower than 3$\times$10$^{21}$~cm$^{-2}$ (blue violet points between 
100$\degr\leq\ell\leq$120$\degr$); 2) Galactic anticentre sources (orange 
points with 150$\degr\lesssim\ell\lesssim$190$\degr$) generally between 3 to 
6$\times$10$^{21}$~cm$^{-2}$, 3) intermediate-$\ell$ sources through the Outer 
Galaxy (90$\degr\leq\ell\leq$150$\degr$; blue-violet-red) with
N$_{\textsc{HI}}=$6 to 9$\times$10$^{21}$~cm$^{-2}$, and 4) low-longitude 
sources with $\ell<$90$\degr$ that pass through the Inner and Outer Galaxy and 
that generally have the highest columns in the sample 
(N$_{\textsc{HI}}\gtrsim$9$\times$10$^{21}$~cm$^{-2}$), though many have 
smaller columns where we look down the lower-density interarm region between 
the Sagittarius and Local arms.

A general analysis of the RM scatter in each column density zone is telling. In 
21~cm CGPS continuum maps no \ion{H}{2} regions are observed in zone~1, and 
given that sources in this region have RM near zero, the strength of the GMF 
here and the $n_e$, relative to the plane regions, are small. Hence the scatter 
in zone~1 RMs is dominated by the small-scale random component of the field. 
Zone~2 is also a region of low RM (since the LOS projection of the large-scale 
GMF is diminished towards the anticentre) but these sources are in the Galactic 
plane and are seen towards many bright catalogued \ion{H}{2} regions; scatter 
in this region would thus be due to both the effects of small-scale variations 
in the GMF and of random fields and high $n_e$ in \ion{H}{2} regions. However, 
the scatter in zone~2 is barely higher than in zone~1, thus the scatter in RM, 
and the mean RM itself in these directions is not dominated by the \ion{H}{2} 
regions seen along these paths. Zone~3 shows moderately increased scatter with 
more true {}``outliers'' on either side of the mean, likely reflecting the 
longer integrated path length which would intercept more \ion{H}{2} regions. 
However, the majority of zone~3 sources are tightly clustered around the mean 
RM$<$0, similar to the concentration in zone~1. Again, we conclude that while 
\ion{H}{2} regions likely do contribute to the scatter in RM, they are not the 
dominant component along the LOS nor across the plane on the sky \cite[as 
mentioned in Sec.~\ref{obs} this conclusion is also demonstrated in][]{brow03}. 
Sources in zone~4 have long lines-of-sight (turquoise-blue points for 
65$\degr\lesssim\ell\lesssim$85$\degr$) that pass down a substantial column of 
the Local Arm and are mainly parallel with the large-scale field in that arm. 
As suggested in \citet{brow01} the small-scale field is likely correlated with 
the large-scale field, and therefore, these lines-of-sight show significant 
variability in $B_{\parallel}$. They also pass through more $n_e$ structures 
that are located within this arm (particularly Cygnus~X), all of which 
contribute to strong spatial variability in RM. Finally for direction 
$\ell\lesssim$65$\degr$ (pure green points) the large-scale field is observed 
to \textit{reverse} direction through the Sagittarius Arm tangent and the 
interarm region beyond this arm \citep[][]{vane11}. The very high scatter in 
zone~4 is therefore likely from all three sources.

\subsection{Binned Results}
\begin{figure*}[!ht]
\centerline{
\includegraphics[scale=1]{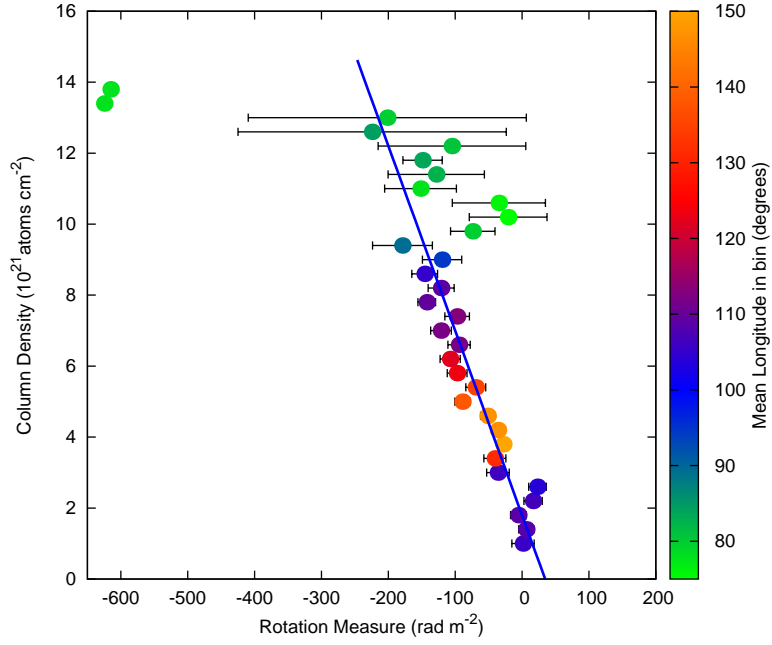}
}
\caption{Sources in Figure~\ref{plot1} binned in column density bins 
4$\times$10$^{20}$~atoms~cm$^{-2}$ wide. The robust mean and error in RM for
each bin is plotted.
}
\label{plot2}
\end{figure*}

%It is apparent in Figure~\ref{plot1} that for a given column density range 
%there is substantial variation in RM observed towards a source. 
To elucidate the underlying relationship, we group the RM data into column 
density bins 4$\times$10$^{20}$~cm$^{-2}$ wide, and calculate the robust mean 
rotation measure in each bin using maximum likelihood estimates 
({}``M''-estimates) to mitigate the effects of outliers (like \ion{H}{2} 
regions). Only bins with $N\geq$3 RM sources are considered, and the 
uncertainty in the mean of each bin ($\Delta$RM) is the median absolute 
deviation over $\sqrt{N}$. Results are shown in Figure~\ref{plot2}. The robust 
M-estimates result in a quite well-determined mean RM for given values of 
N$_{\textsc{HI}}$. %One 
%source of the scatter could be a random variation in the ionization fraction 
%(ratio of electrons-to-atoms) in the ISM (such a variation might be expected), 
%but we also see RMs on either side of zero in each bin, suggesting that a 
%random component to the Galactic magnetic field (GMF) also plays a significant 
%role in creating the scatter. 
A robust least-squares fit to the binned data weighted by $\Delta$RM shows an 
underlying linear relationship with a slope of
$-$19.2$\pm$1.6~rad~m$^{-2}$/10$^{21}$~atoms~cm$^{-2}$, and at zero column 
density a constant of 34.5$\pm$8.5~rad~m$^{-2}$. The non-zero column density 
here at RM$=$0 is likely a natural consequence of lines-of-sight through 
regions with $B_{\parallel}=$0, such as in directions towards the anticentre. 
The Pearson (correlation) coefficient of the points after binning is r=0.78.

\section{Discussion of Results}\label{res}
The strong linear correlation between the mean RM and N$_{\textsc{HI}}$ could 
arise because each depends on something else in common. For example, lower 
rotation measures and column densities are both found in directions where the 
integrated path length is smaller through the disk; i.e. high longitudes (near 
the anticentre) or high latitudes. Directions near $\ell=$180$\degr$ are also 
where there would be no LOS-component of a purely azimuthal field, since 
geometrically $B_{\parallel}=\left|\vec{B}\right|~R_0/R~\textrm{sin}~\ell$, and 
for a circularly oriented field of uniform strength 
$B_{\parallel}\propto\textrm{sin}~\ell$ along a circle of radius $R$. 
\textit{Along} the LOS through the Outer Galaxy the uniform-field's strength 
and the electron density fall with $R^{-1}$ \citep[see][]{brow03b} whereas the 
density of \ion{H}{1} stays nearly constant to at least $R=$13~kpc \citep[see 
Fig.7.15~][]{burt88}, falling thereafter with an exponential decay 
\citep[e.g.][]{fost06}. In high latitude directions the atomic and electron 
densities and the field strength also diminish, each with different 
characteristic heights \citep[e.g.][]{dick90,gaen08,kron11}, but the amount of 
disk matter sampled along each direction changes in a more complicated way than 
it would across longitude-varying directions at constant latitude due to the 
warping of the midplane and the flaring of its thickness \citep[][]{fost06}. 

\begin{figure*}[!ht]
%\centerline{
\hspace{-2.5cm}
\includegraphics[scale=1]{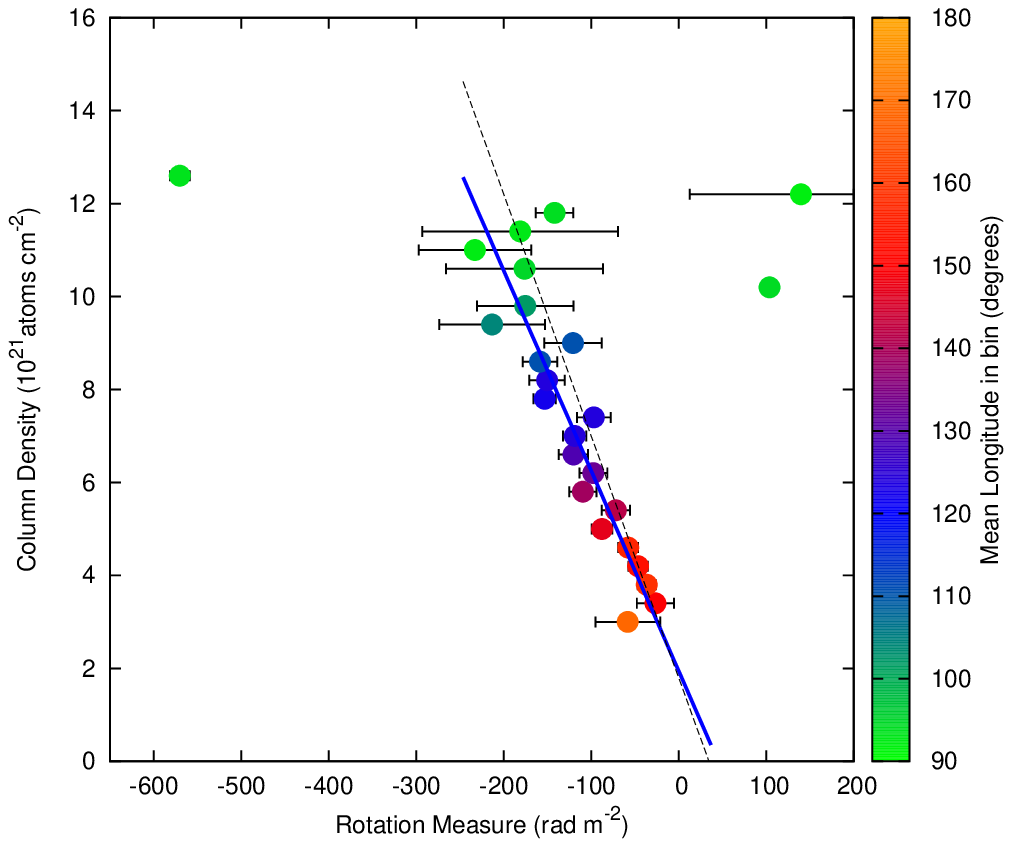}\hspace{-2.5cm}
\includegraphics[scale=1]{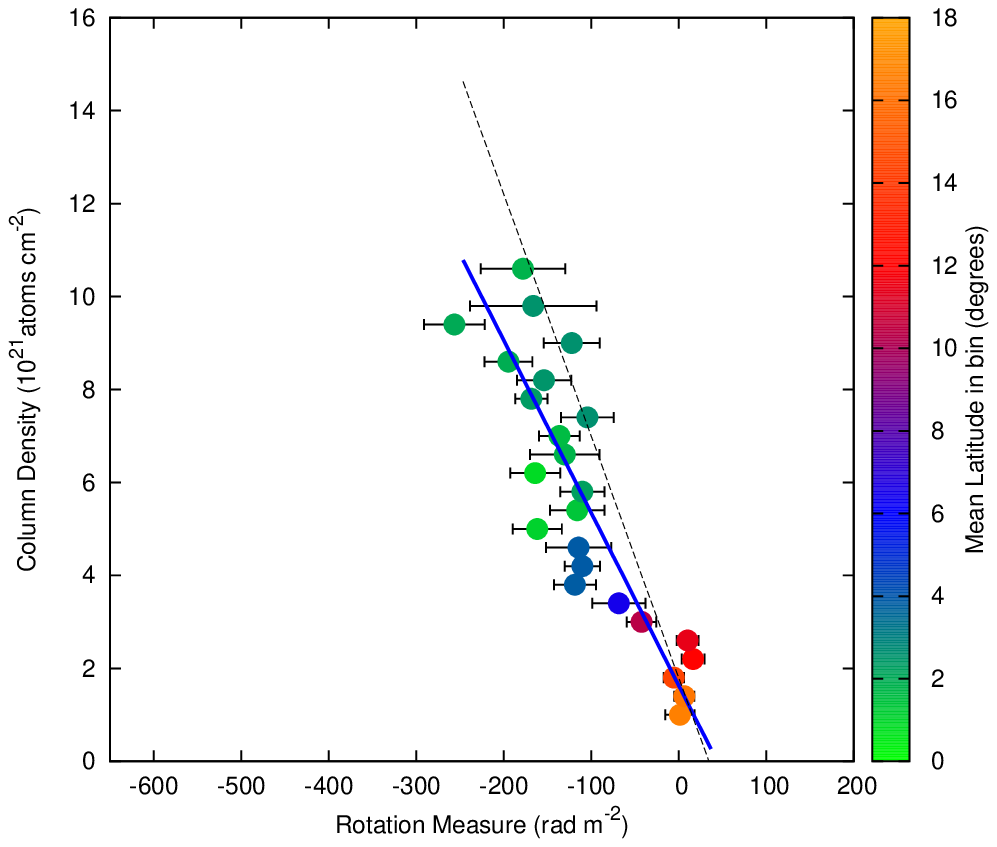}
%{f01b.eps}
%}
\caption{(Left) Region~1 described in the text, Sec.~\ref{res}. (Right) 
Region~2. The dashed line plotted in each is the one fitted to the entire
set of RMs, shown in Figure~\ref{plot2}.}
\label{plot3}
\end{figure*}

To test this possibility, we isolate two regions of the CGPS. In region~1 we 
restrict the dependence to longitude only by looking at sources across a wide 
longitude strip 90$\degr\leq\ell\leq$180$\degr$ in a narrow latitude 
range centred on the plane ($-$3.5$\degr\leq b\leq+$5$\degr$; 1110 sources). 
For region~2 we take sources in a vertical strip across a wide latitude 
($-$3.5$\degr\leq b\leq+$18$\degr$) centred within a narrow longitude range 
(95$\degr\leq\ell\leq$125$\degr$; 692 sources). The dependence of 
integrated path length on direction across each strip should be markedly 
different, since it is expected that $n_e,~n_{\textsc{HI}}$ and $B_{\parallel}$ 
are changing in very different ways and for different reasons across each 
region. If there is an underlying dependence on direction the observed 
relationship between RM and N$_{\textsc{HI}}$ in each should also be different. 
Remarkably, however, the fitted lines in region~1 (slope 
$-$23.2$\pm$2.3, intercept 45.0$\pm$13.8, r=0.87) and region~2 
($-$26.9$\pm$2.5~rad~m$^{-2}$/10$^{21}$~atoms~cm$^{-2}$, 
44.0$\pm$10.9~rad~m$^{-2}$; r=0.79) are essentially identical 
(Figure~\ref{plot3}), evidence that the ratio of RM/N$_{\textsc{HI}}$ is 
independent of direction $\ell$ and of direction $b$.

A linear relationship observed between these variables would still exist 
whether we correct each for the same direction/path effect or not, with only 
the slope potentially changed. The remarkable constancy of the slope in 
different regions of the survey is thus indicating that something other than 
mutual directional dependence is being observed.

\section{Magnetic Field and Ionization Fraction Estimates}
\label{B-est}
We can use Equation~\ref{ratio} to estimate the bulk mean GMF strength in the 
CGPS region for a given mean ionization fraction, and vice-versa. For this we 
must be mindful that if there are reversals along the LOS, the value of 
the ratio RM/N$_{\textsc{HI}}$ will be closer to zero and calculated values for  
$\left<B_{\parallel}\right>$ or $\left<x_e\right>$ will be lower limits to the 
actual values. For our estimate we avoid RMs in the Inner Galaxy Quadrant I 
(to avoid the field reversal), and restrict our estimate to QII 
(90$\degr\leq\ell\leq$180$\degr$) where the uniform component will be pointed 
away from the observer and RMs tracing the field will be negative. Other 
smaller-scale contributors to RM (\ion{H}{2} regions, for example) are expected 
to have an unsubstantial impact on the mean since the large number of 
lines-of-sight we use ensures that localized regions do not dominate the 
large-scale trends in each bin \citep[again, see][]{brow03}. A ratio 
RM/N$_{\textsc{HI}}=-$21.8$\pm$1.5~rad~m$^{-2}$/10$^{21}$~atoms~cm$^{-2}$ is 
found, with r=0.94.

Within a few hundred parsecs of the Sun, \citet{jenk13} find $x_e\sim$0.08 (for 
a mean total H density of $n_{\textsc{H}}=$0.5~cm$^{-3}$). If we take this as 
representative of the bulk volume-averaged value, then the LOS component of
the Outer Galaxy GMF is $\left<B_{\parallel}\right>\sim-$0.95$\pm$0.06~$\mu$G. 
A slightly higher mean field of $-$1.0$\pm$0.1~$\mu$G is found if we also 
restrict sources to the immediate Galactic plane; $\left|b\right|\leq$5$\degr$ 
(region~1 in Sec.~\ref{res}; see Figure~\ref{plot3}, 
RM/N$_{\textsc{HI}}=-$23.2$\pm$2.3~rad~m$^{-2}$/10$^{21}$~atoms~cm$^{-2}$, 
intercept 45.0$\pm$13.8~rad~m$^{-2}$). Both are very consistent with 
$\left<B\right>=-$1.2$\pm$0.48~$\mu$G, the value near the Sun from 
\citet{vane11} and an upper-limit for the mean LOS-projection component 
(\citeauthor{vane11} use the same RMs as we do but rely on the $n_e$ model 
of \citet{cord02}).

Alternatively we can use Eqn.~\ref{ratio} to crudely estimate a lower 
limit to the ionization fraction in the Galactic plane, QII, given a mean field 
strength estimate for the ISM \citep[$\left<B\right>\simeq-$1.2$\pm$0.48~$\mu$G 
in][again as an upper-limit to $\left<B_{\parallel}\right>$]{vane11}. With 
this, $x_e\gtrsim$6.8\%, which is above the WNM phase calculation of 
$\simeq$2\% by \citet{wolf95} (for $n_{\textsc{H}}=$0.4~cm$^{-3}$) and more 
consistent with the \citeauthor{jenk13} result for the local ISM of 8\%. If we 
use the local regular magnetic field strength estimate 
$\left<B\right>\simeq-$2~$\mu$G of \citet{sun08}, then $x_e\gtrsim$4.2\%, 
intermediate between the two previously published results. %The ionization fraction is related to the ionization rate 
%in the WNM by $x_e\sim$0.03$\sqrt{\zeta_{\textsc{TOT}}/\left(10^{-16}\right)}$ 
%\citep[from][valid for canonical WNM values of $T=$8000~K and 
%$n_{\textsc{H}}=$0.4~cm$^{-3}$]{kulk88}, where the total rate 
%$\zeta_{\textsc{TOT}}$ is the sum of ionization rates by cosmic rays and soft 
%X-rays. Our measured slope RM/N$_{\textsc{HI}}$ for the plane-region and the 
%value for $\left<B_{\parallel}\right>$ from \citet{vane11} suggests that on 
%average $\left<\zeta_{\textsc{TOT}}\right>\sim$4.4$\times$10$^{-16}$~s$^{-1}$ 
%per H atom in QII. The fraction of cosmic-ray to X-ray ionization rates 
%$\zeta_{\textsc{CR}}/\zeta_{\textsc{XR}}$ varies with column density 
%\citep[e.g. Fig.~2 in][]{wolf95}, so a measurement of the volume-averaged total 
%ionization rate $\left<\zeta_{\textsc{TOT}}\right>$ is all we can provide in 
%this paper.
In any case, using our measured slope RM/N$_{\textsc{HI}}$ in Eqn.~\ref{ratio} 
with widely-accepted values for the uniform field strength results in low 
($<$10\%) ionization fractions that are consistent with the range accepted for 
the WNM. 

\section{Discussion and Summary}
The consistency of the field estimated from Eqn.~\ref{ratio} (using WNM values 
for $x_e$) with the field estimated by \citet{vane11} \citep[using dispersion 
measures from the $n_e$ model of][]{cord02} suggests that most of the electrons 
in the large-scale MIM responsible for pulsar DM are related to the atoms in 
the WNM. Our observed ratio RM/N$_{\textsc{HI}}$ seems to be sufficient to 
predict the magnitude of the magnetic field in the general MIM, suggesting that 
electrons from fully ionized regions of the WIM account for very little RM and 
that a very large fraction of the electrons seen in the Galactic plane arise 
from the partially ionized WNM, compared to the fully ionized, low density WIM. 
However, as the field and ionization fractions of the WNM are not known with 
terrific accuracy, we cannot provide a viable quantitative estimate of this 
fraction. Nonetheless, qualitatively it appears that the boundary between the 
WNM and MIM is very much indistinct and that the GMF itself may be supported by 
the partially ionized WNM. 

%Finally, it is to be noted that Eqn.~\ref{ratio} is as valid for shorter
%regions ({}``pockets'') of the ISM as it is for longer paths, since 
%RM/N$_{\textsc{HI}}$ is independent of length. Therefore measurement of this
%ratio and Eqn.~\ref{ratio} may offer a way of studying either the (unknown) 
%magnetic field or ionization fraction in smaller discrete regions of the ISM
%where the other (known) parameter can be more safely assumed constant 
%throughout.

Using observations from the CGPS we have found a linear correlation between 
rotation measure and \ion{H}{1} column density (for the optically thin WNM 
gas). The slope RM/N$_{\textsc{HI}}$ observed is of the same order expected 
from a simple theoretical calculation (Eqn.~\ref{ratio}) with current estimates 
of the ionization fraction in the WNM and the bulk LOS magnetic field in the 
Outer Galaxy. Since the ratio RM/N$_{\textsc{HI}}$ ultimately has units of 
radians of rotation per 10$^{21}$ H atoms it is independent of path length, and 
its observed value can be used to estimate mean magnetic fields and/or 
ionization fractions in any region of interest, regardless of its dimension. 
Therefore measurement of this ratio and Eqn.~\ref{ratio} may offer a way of 
studying either the (unknown) magnetic field or ionization fraction in smaller 
discrete regions of the ISM where the other (known) parameter can be more 
safely assumed constant throughout. This may offer some advantage over the 
current approach of estimating field strengths with the (path-integrated) 
dispersion measure.

\acknowledgements
The authors would like to thank the referee for their careful and thoughtful
read of our manuscript, and constructive suggestions to improve it. TF would 
like to thank Dr. Sean Dougherty, Director of the Dominion Radio Astrophysical 
Observatory (DRAO) for his support and hospitality during the author's 
sabbatical stay. This work has been supported partially via a grant from the 
Brandon University Research Committee (BURC) to TF. DRAO is a National Facility 
operated by the National Research Council. The Canadian Galactic Plane Survey 
is a Canadian project with international partners, and has been supported by 
the Natural Sciences and Engineering Research Council (NSERC).

%\bibliographystyle{aa}
%\bibliography{ms}

\end{document}